\documentclass[iop]{emulateapj}
\usepackage{hyperref}

\shorttitle{Photometry as a proxy for stellar activity in radial velocity analyses}
\shortauthors{Kosiarek \& Crossfield}

\begin{document}


\title{Photometry as a proxy for stellar activity in radial velocity analyses}

\author{Molly R.\ Kosiarek\altaffilmark{1,$\dagger$,*}}
\author{Ian J.M. Crossfield\altaffilmark{2}}

\altaffiltext{1}{Department of Astronomy and Astrophysics, University of California, Santa Cruz, CA 95064, USA}
\altaffiltext{$\dagger$}{NSF Graduate Research Fellow}
\altaffiltext{*}{molly.kosiarek@gmail.com}
\altaffiltext{2}{Kansas University Department of Physics and Astronomy, 1082 Malott, 1251 Wescoe Hall Dr. Lawrence, KS 66045}

\begin{abstract} 
Stellar activity remains a limiting factor in measuring precise planet parameters from radial velocity spectroscopy, not least in the search for Earth mass planets orbiting in the habitable zones of Sun-like stars. 
One approach to mitigate stellar activity is to use combined analyses of both radial velocity and time-series photometry.
We present an analysis of simultaneous disk-integrated photometry and radial velocity data of the Sun in order to determine the useful limits of a combined analysis.
We find that simple periodogram or autocorrelation analysis of solar photometry give the correct rotation period $<$50\% of the time. We therefore use a Gaussian process to investigate the time variability of solar photometry and to directly compare simultaneous photometry with radial velocity data. 
We find that the hyperparameter posteriors are relatively stable over 70 years of solar photometry and the amplitude tracks the solar cycle. 
We observe good agreement between the hyperparameter posteriors for the simultaneous photometry and radial velocity data.
Our primary conclusion is a recommendation to include an additional prior in Gaussian process fits to constrain the evolutionary timescale to be greater than the recurrence timescale (ie., the rotation period) to recover more physically plausible and useful results.
Our results indicate that such simultaneous monitoring may be a useful tool in enhancing the precision of radial velocity surveys.

\end{abstract}



\section{Introduction}

\begin{figure*}[tbp]
\includegraphics[width=0.99\textwidth]{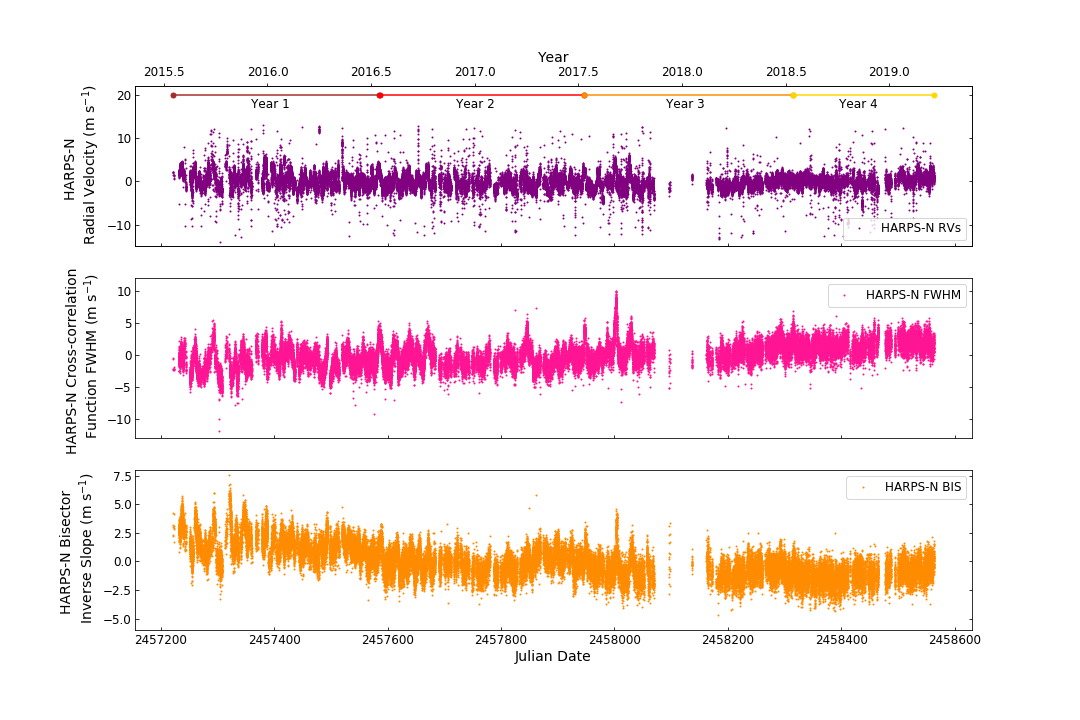}
\caption{HARPS-N solar radial velocity data (top, purple) corrected for barycentric motion and with all solar system planets removed, HARPS-N full-width half-maxiumum of the cross-correlation function (middle, pink), and the bisector inverse slope (bottom, orange) from \citet{CollierCameron2019}. The four years of HARPS-N data are labeled as specified in the rest of the analysis. \label{fig:harpsn}}
\end{figure*}

Transit surveys are detecting hundreds of Earth-sized planets and measuring their sizes and orbital properties\footnote{NASA Exoplanet Archive: \url{https://exoplanetarchive.ipac.caltech.edu/}}. To understand the composition and potential habitability of these planets, mass measurements are needed to calculate a bulk density and interpret future atmospheric transmission spectroscopy measurements \citep{Batalha2019}. Furthermore, future high-contrast characterization of Earth-mass planets in the habitable zone of nearby stars would benefit from target identification by precise radial velocity surveys \citep{Gaudi2020}. 
These Earth-like planets orbiting in the habitable zones of G stars produce a radial velocity signal of only 10 cm s$^{-1}$. Due to the high precision and long monitoring time needed, no true Earth analogues currently have mass measurements from the radial velocity method. 

Radial velocity instrument stability and calibration is rapidly approaching the ability to detect an Earth-like signal. For example, NEID has an error budget of 27 cm s$^{-1}$ \citep{Halverson2016}, ESPRESSO is achieving a 28 cm s$^{-1}$ dispersion on sky over a single night \citep{Pepe2014,Pepe_EPRV}, and laser frequency comb measurements on EXPRES are showing an instrumental precision of $<$10 cm s$^{-1}$ \citep{Zhao_EPRV,Blackman2020,Petersburg2020}. Yet there is much work needed to mitigate stellar activity to detect such a small signal on sky. 

The HARPS-N team have been collecting disk-integrated radial velocity observations of our Sun over the last four years \citep{CollierCameron2019}. After accounting for the radial velocity shifts from all of the solar system planets and thoroughly vetting for data quality, there remains an underlying solar variability signal of 5 m s$^{-1}$ with a daily RMS scatter of $<$1 m s$^{-1}$. Stellar activity therefore remains the largest ``noise'' component in radial velocity analyses of the Sun, and will likely limit future surveys unless this noise can be mitigated.

Stellar activity associated with a star's rotation period can affect the analysis of orbiting planets or be mistaken as a planetary signal due to their overlapping timeframes of days to tens of days \citep[eg.][]{Haywood2018,Robertson2014,Mortier2017}.
Starspots cause variations in stellar line profiles and centroids (eg. Vogt, Penrod, \& Hates 1987); therefore monitoring stellar rotation with photometry may be a valuable tool for identifying and mitigating these stellar activity signals in radial velocity data. 
Previous works have found similar periodicities in photometry and radial velocity data
and have used this correspondence to improve the precision of the planet parameters \citep[eg.][]{Aigrain2012,Haywood2014,LopezMorales2016,Kosiarek2019}. 

In this paper, we explore the relationship between Gaussian process parameters derived from solar photometry to those derived from solar radial velocity data in order to better understand how photometry can be used for activity mitigation. We describe the data used in this paper and look for common periodicities between the datasets in Section 2. We introduce Gaussian processes and our analysis methods in Section 3. We examine the time variability of solar photometry in Section 4.1, followed by a direct comparison between Gaussian process parameters derived from solar photometry and radial velocity data in Section 4.2 before concluding with advice for future observations in Section 5.

\section{Solar Datasets}

The Sun makes a particularly good test case due to the abundance and precision of solar monitoring. In this work, we examine 1) the time variability of solar photometry over 70 years of data and 2) the relationship between photometry and radial velocity data through comparing four years of simultaneous solar photometry and radial velocity data. 

The HARPS-N team recently published a large solar radial velocity dataset taken with a solar telescope that feeds disk-integrated sunlight to the HARPS-N spectrograph \citep{CollierCameron2019}. The radial velocity data span nearly four years, from July 2015 to March 2019 (Figure~\ref{fig:harpsn}). Dozens of datapoints are taken per day, weather permitting, with 5 minute integrations and result in a typical precision of 0.43 m s$^{-1}$.  

The HARPS-N data reduction package also produces two line measurements alongside the radial velocity data, the full-width half-maximum (FWHM) of the cross-correlation function (CCF) and a measurement of the asymmetry of the CCF called the bisector inverse slope (BIS). These two measurements can be used as stellar activity indicators, therefore we will compare them alongside the radial velocity data throughout our analysis.

\begin{figure*}[htp]
\includegraphics[width=0.95\textwidth]{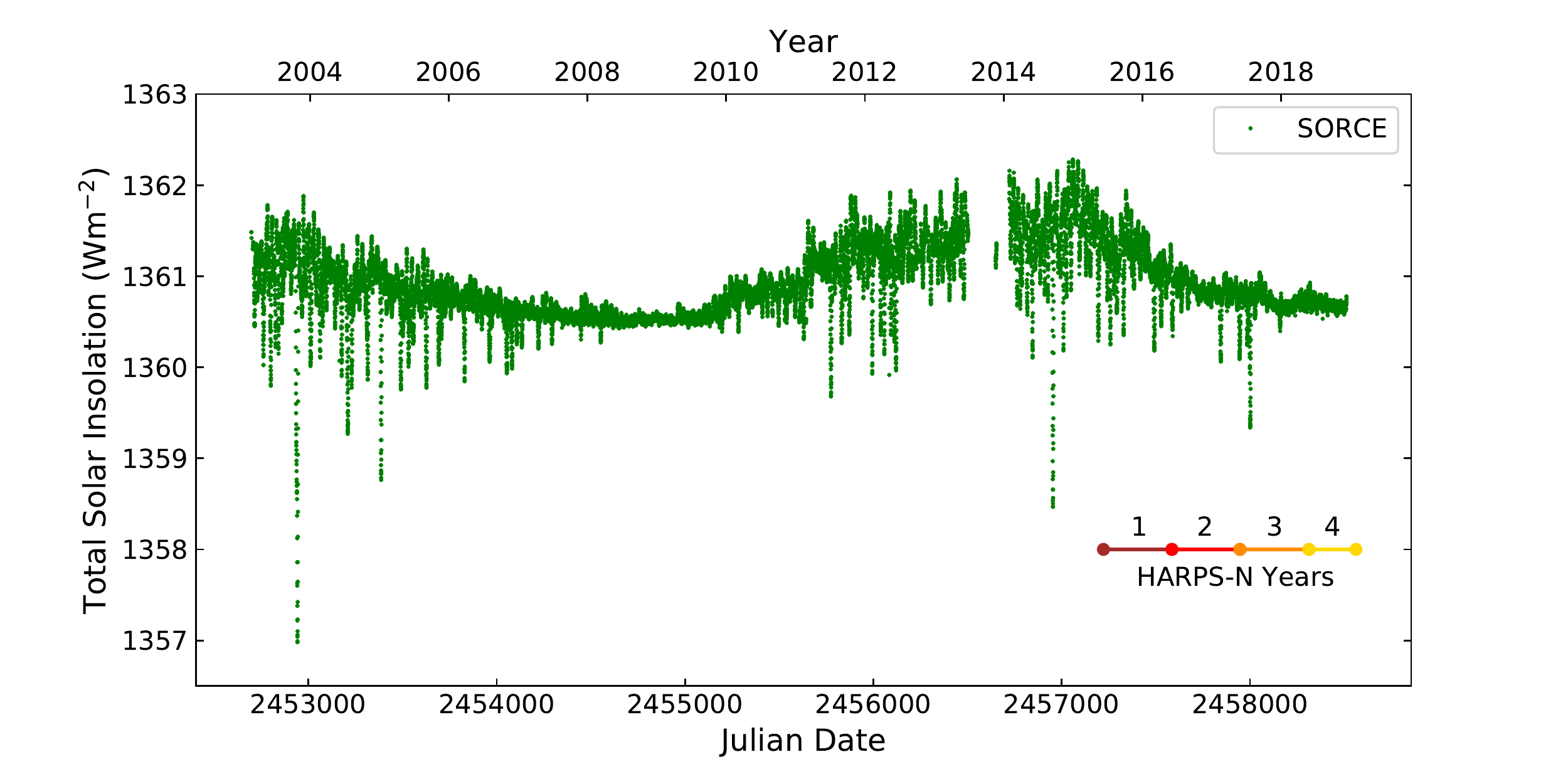}
\caption{SORCE total solar insolation data from February 2003 to August 2019. The four years of data that overlap with the HARPS-N dataset are labeled.  \label{fig:sorce}}
\end{figure*}

The SOlar Radiation \& Climate Experiment (SORCE) measures the total solar irradiance (TSI) with the total irradiance monitor \citep{Lawrence2000}. The TSI dataproducts\footnote{\url{http://lasp.colorado.edu/home/sorce/data/tsi-data/}} include daily and 6-hour average irradiances normalized to a distance of 1 AU and the data have a typical precision of 0.5 W m$^{-2}$ (Figure~\ref{fig:sorce}). 

\begin{figure*}[htp]
\includegraphics[width=0.95\textwidth]{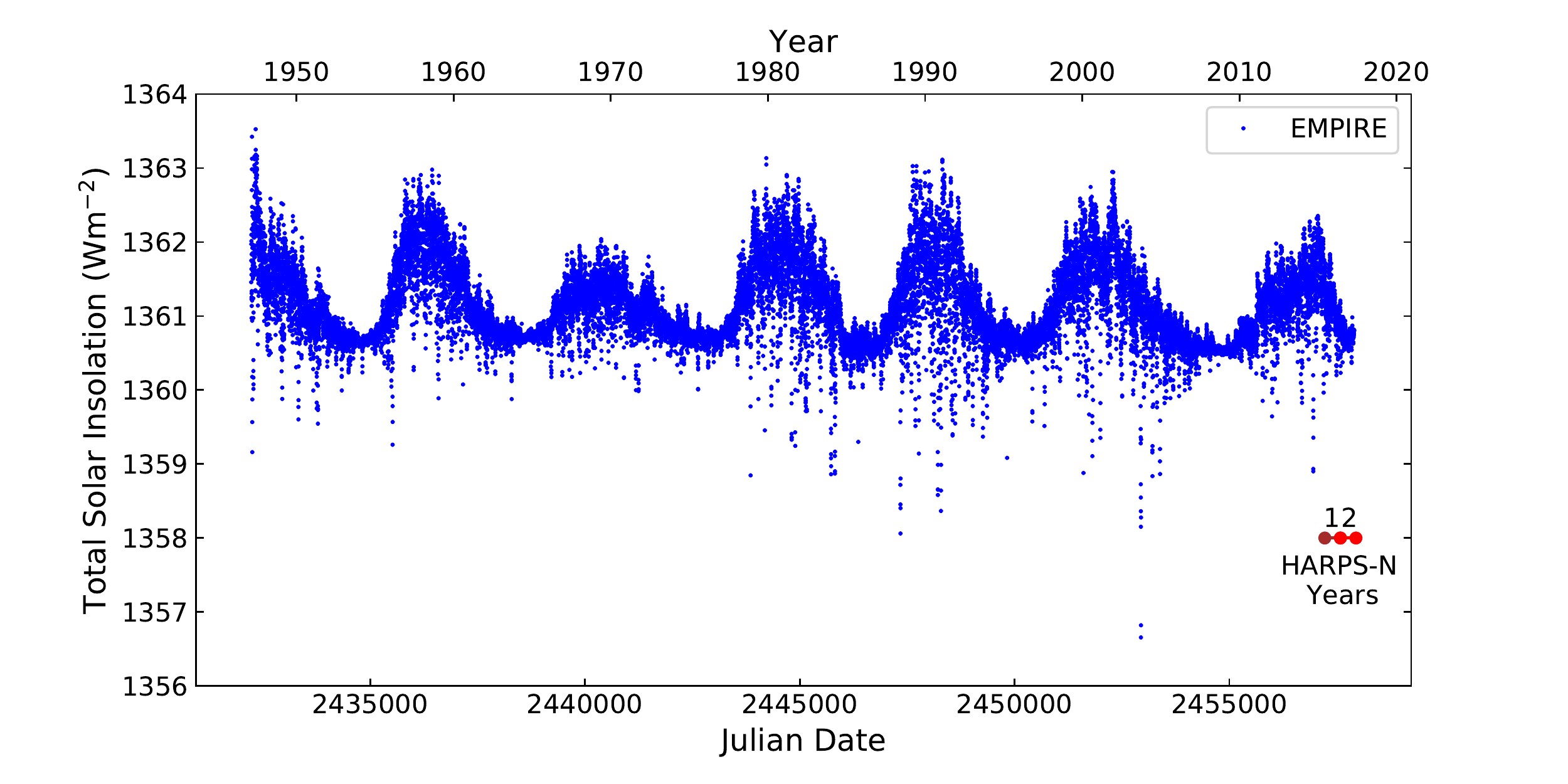}
\caption{EMPIRE total solar insolation data from February 1947 to September 2016. The two years of data that overlap with the HARPS-N dataset are labeled. \label{fig:empire}}
\end{figure*}

The EMPirical Irradiance REconstruction (EMPIRE) is a solar irradiance model with the goal of providing uninterrupted and coherent TSI time series for climate modeling \citep{Yeo2017}. The solar irradiance is calculated by a linear combination of solar activity indices connected to sunspots and faculae. The dataset begins February 1947 and extends to September 2016 (Figure~\ref{fig:empire}). EMPIRE overlaps with the SORCE dataset from 2003-2016 with good agreement (RMS difference of 0.12 Wm$^{-2}$). Therefore, this work will use the EMPIRE dataset when discussing variations over time due to its much longer baseline and the SORCE dataset when comparing with the HARPS-N radial velocity data due to the overlap between these two datasets. 

\subsection{Initial Data Comparisons}

To directly compare the EMPIRE and SORCE photometry with the HARPS-N radial velocity, we first split each of the datasets into year-long segments that overlap with the timescale of the HARPS-N data. These segments are labeled ``Year 1-4'' in Figures 1--3. 

The three datasets used in this project have different sampling cadences and distribution. To normalize the inputs for each fit, we binned the datapoints in daily bins with uncertainties that represent the standard deviation of the points. This binning was also performed to focus on the solar rotation timescale, as opposed to short timescale activity such as p-modes and granulation. Binning on a daily cadence is also standard practice in many precise radial velocity analyses \citep{Dumusque2011, Chaplin2019}. 

\begin{figure}[htp]
\includegraphics[trim=0 80 0 0,clip, width=0.5\textwidth]{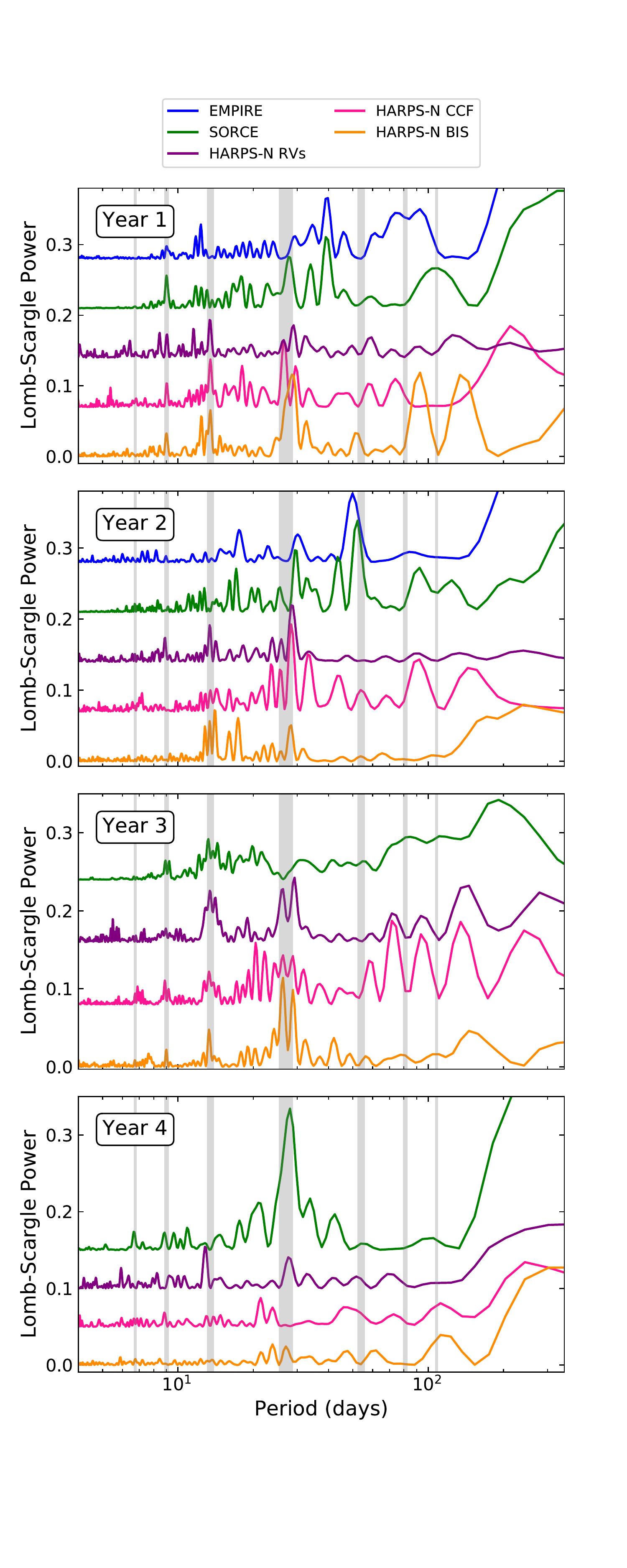}
\caption{Periodogram comparison of the solar photometry and radial velocity data. All datasets are plotted with individual y-offsets for clarity. The stellar rotation period (27d, thick grey line) and its harmonics (thin grey lines) are plotted for comparison. We find that many of the peaks in all datasets line up with the solar rotation period and its harmonics. \label{fig:periodogram}}
\end{figure}

We initially looked for common periodicities in the datasets using two different techniques, a Lomb-Scargle periodogram and autocorrelation.  The Lomb-Scargle periodogram results are shown in Figure~\ref{fig:periodogram}. The majority of the peaks occur at the solar rotation period or at its harmonics. In all four years, the HARPS-N RV data have peaks at the stellar rotation period (27 days) and the 1/2 and 1/3 harmonic. In two years, Year 1 and Year 4, the peak at 1/2 of the rotation period is the highest and the peak at the rotation period is the second highest. The HARPS-N FWHM and BIS data primarily follow the radial velocity data, except for Year 4 which has few significant peaks. For the photometry, the majority of the peaks occur at the stellar rotation period or its harmonics, however the peaks are less consistent than the HARPS-N RV data. 

\begin{figure}[htp]
\includegraphics[trim=0 80 0 0,clip, width=0.5\textwidth]{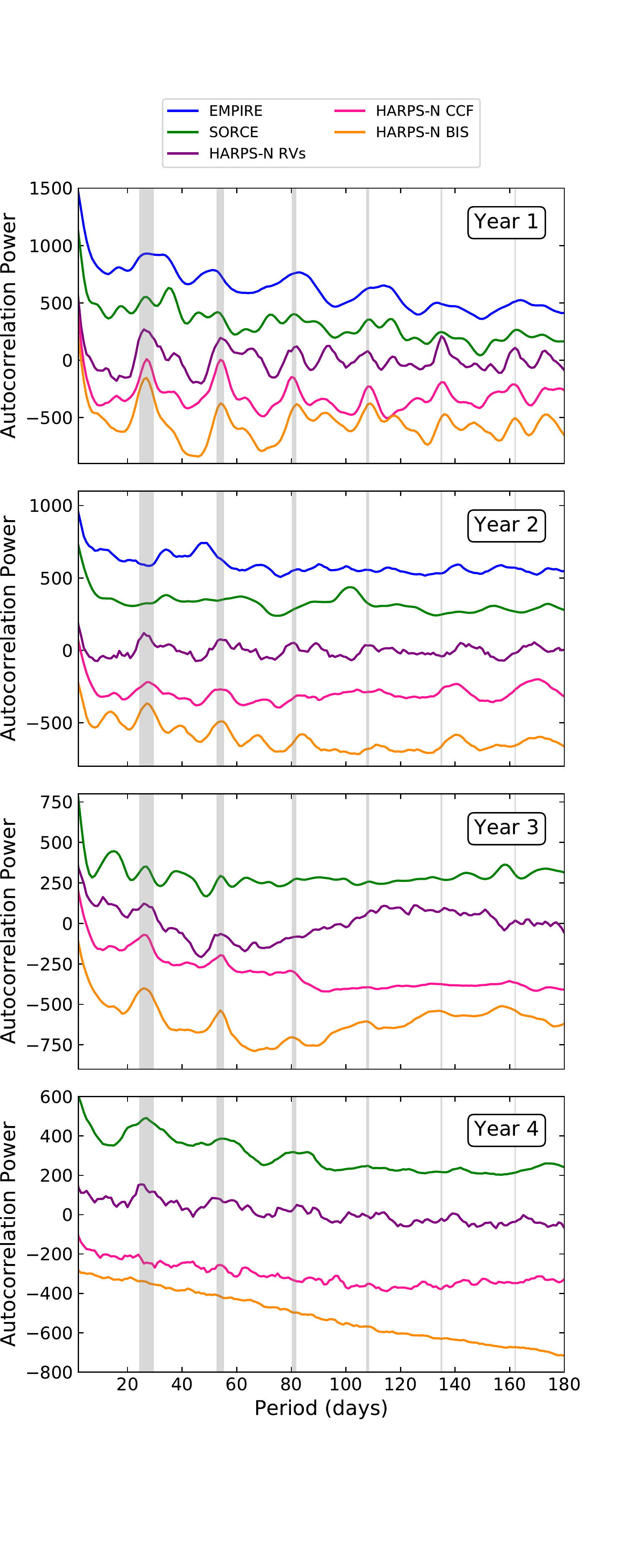}
\caption{Autocorrelation comparison of the solar photometry and radial velocity datasets. The photometry (SORCE \& EMPIRE) and line indicators (HARPS-N FWHM \& HARPS-N BIS) are normalized to the scale of the HARPS-N RVs and are plotted with a y-offset for clarity. The stellar rotation period (27d, thick grey line) and its harmonics (thin grey lines) are plotted for comparison. Many of the peaks in both datasets line up with the stellar rotation period and its harmonics. \label{fig:autocorrelation}}
\end{figure}

Due to the stochastic nature of stellar activity, the highest peak in a periodogram is often not at the stellar rotation period 
\citep{Boisse2011,Nava2019arxiv}, therefore we also examine autocorrelation plots for all of our datasets. 
We first linearly interpolate the HARPS-N data to a uniform daily cadence to perform the autocorrelation using numpy.interp and numpy.correlate in Python. 
The autocorrelation for the photometry and radial velocity data over Years 1-4 are shown in Figure~\ref{fig:autocorrelation}. The HARPS-N RV autocorrelation has a distinct ``sawtooth" pattern in Years 1, 2, and 4, with peaks at the stellar rotation period and multiples thereof. Year 3 has a break in the middle of the dataset that likely creates the broad peak at 125d and partially washes out the stellar rotation signal. The FWHM and BIS also peak at the solar rotation period and its harmonics in Years 1-3; there are no significant peaks in Year 4. The photometry follows the same ``sawtooth" pattern in Years 1 and 4. 
Years 2 and 3 have larger variance in the amplitude of the total solar insolation which may contribute to the inconsistent peaks. In summary, there is good agreement between the SORCE, EMPIRE, and RV data for Years 1 and 4, and good agreement between the RV, FWHM, and BIS data for Years 1--3.  

\subsection{EMPIRE Data Periodicities}

\begin{figure}[htp]
\includegraphics[width=0.48\textwidth]{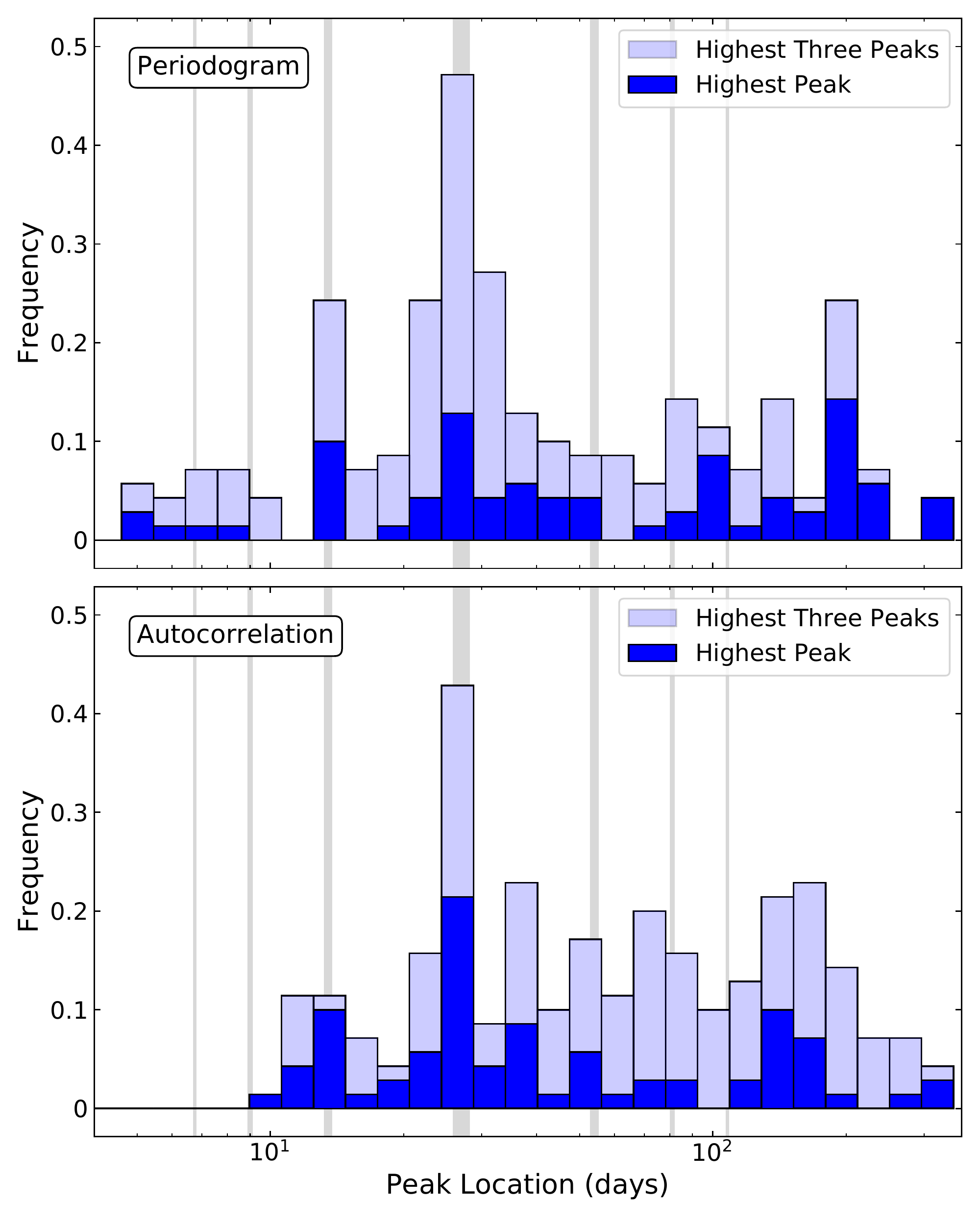}
\caption{Histograms showing the location of the highest (dark blue) and highest three (light blue) peaks in a periodogram (top) and autocorrelation (bottom) analysis of the 70 years of EMPIRE data.
The stellar rotation period (27d, thick greyline) and its harmonics (thin grey lines) are plotted for comparison.
Both histograms show a plurality of peaks at the solar rotation period; however, one should exercise caution when using either of these methods to determine a stellar rotation period as there are a significant number of peaks unrelated to the solar rotation period. 
\label{fig:empirehist}}
\end{figure}

To further examine the accuracy of peridogram and autocorrelation analyses we perform both on each year of the 70 yr EMPIRE dataset. We record the highest three peaks in the periodogram and autocorrelation plots for each year to determine how often the top three peaks are consistent with the solar rotation period, shown as a histogram in Figure~\ref{fig:empirehist}. 
In a periodogram, the solar rotation periods is consistent with the highest peak 14.3\% of the time and one of the highest three peaks 48.6\% of the time. For the autocorrelation, the rotation period is consistent with the highest peak 21.4\% of the time and one of the highest three peaks 44.3\% of the time. As the highest peaks are often at other values unrelated to the solar rotation period, one should exercise caution when using either of these methods to determine a stellar rotation period. 

\section{Methods}

\subsection{Introduction to Gaussian Processes}

In this paper we investigate the validity of using photometry to constrain the hyper-parameter values in a Gaussian process analysis using solar data as a test case. 

Gaussian processes (GP) are a statistical method for modeling correlated noise. Gaussian process regression allows us to determine posterior distributions with uncertainties that reflect effects of stellar activity at specified timescales through a covariance matrix without directly parameterizing functions \citep[e.g.][]{Haywood2014,Grunblatt2015,LopezMorales2016}. 

Stellar rotation activity signals are stochastic: they often match the timescale of planet orbital periods, are quasi periodic due to a combination of periodic stellar rotation and evolving active regions, and are characterized by some degree of smoothness since the active regions do not change instantaneously. 
These stellar signals should be well described by a Gaussian process with a periodic component for the stellar rotation, a component to allow for increasing and decreasing active regions, and a degree of smoothness \citep[eg.][]{Angus2018}. In some cases, radial velocity data are independently able to constrain both the stellar activity and planet parameters \citep{Damasso2017,Faria2016}. 
However, radial velocity data are often too sparse to well constrain Gaussian process hyper-parameters in addition to all of the planet parameters. Therefore, other data sources are used to constrain the hyper-parameters 
and incorporated into the radial velocity analysis as priors \citep[e.g.][]{Haywood2014,Rajpaul2015,Kosiarek2019}.

\subsection{Fitting a Gaussian Process}

\begin{figure*}[htp]
\includegraphics[trim=0 370 30 0,clip,width=0.95\textwidth]{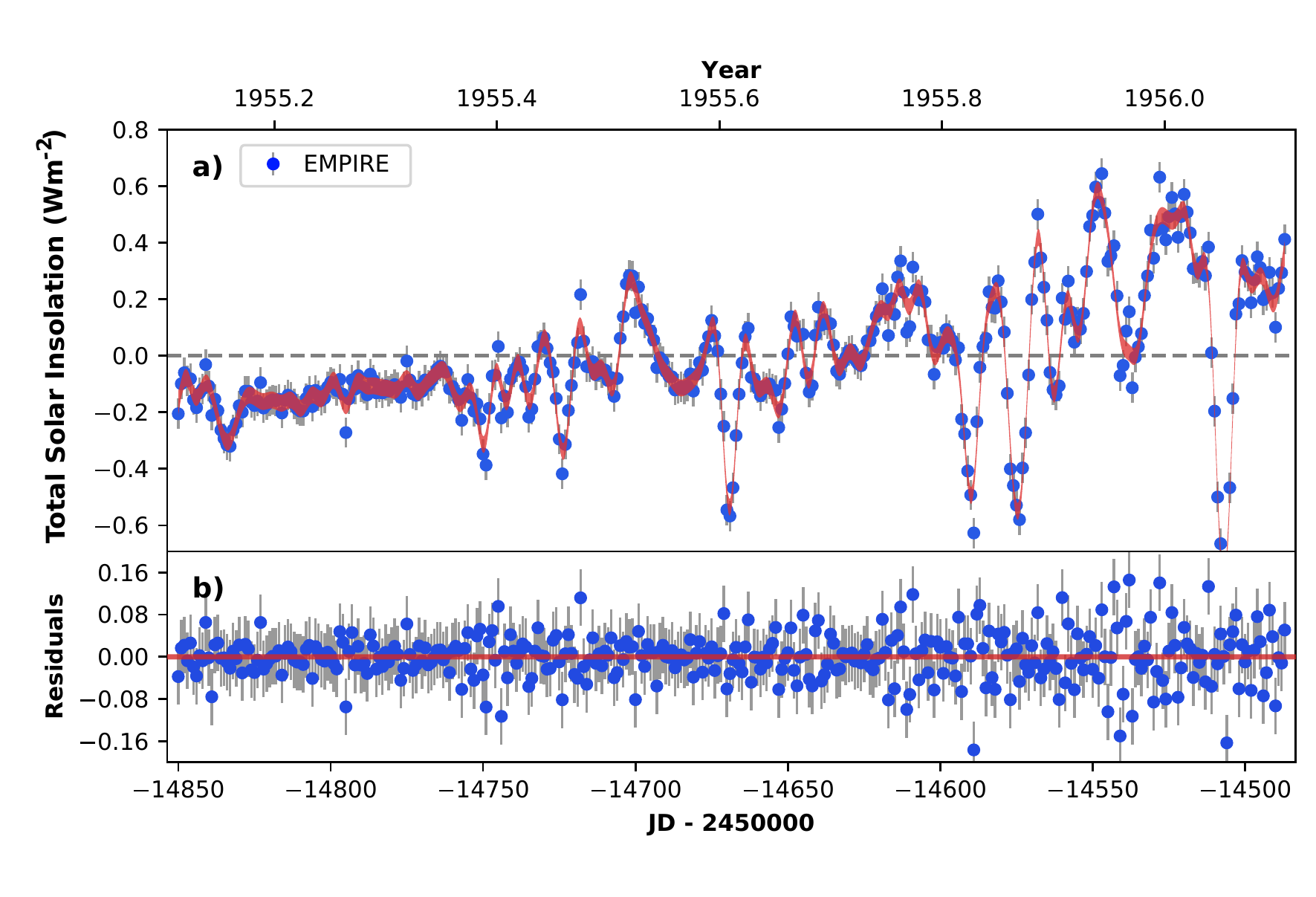}
\caption{Example Gaussian process fit (red line) to one year of EMPIRE total solar irradiance data (blue points). The top panel shows the mean-subtracted data and the bottom panel shows the residuals. A Gaussian process with a quasi-periodic kernel well describes this year of solar photometry.}
\label{fig:empirefit}
\end{figure*}

We model each of our solar datasets using a quasi-periodic GP with a covariance kernel of the form, 

\begin{equation}
\label{eq:kernel}
    k(t,t') = \eta_1^2 \ \rm{exp} \left[-\frac{(\textit{t}-\textit{t'})^2}{\eta_2^2}-\frac{sin^2(\frac{\pi(\it{t}-\it{t'})}{\eta_3})}{\eta_4^2})\right],
\end{equation}
where the hyper-parameter $\eta_1$ is the amplitude of the covariance function, $\eta_2$ is the active region evolutionary timescale, $\eta_3$ is the period of the correlated signal or recurrence timescale, and $\eta_4$ is the lengthscale of the periodic component. This kernel allows for active region evolution through the decay term and a periodic component such as stellar rotation; therefore, it is a suitable kernel choice for fitting stellar activity \citep[eg.][]{Haywood2014,Kosiarek2019}. 

We implement the GP fit using RadVel\footnote{RadVel is available at \url{https://github.com/California-Planet-Search/radvel}} \citep{Fulton2018}. RadVel is an open source Python package that is typically used for fitting radial velocity data with Keplerian orbits. We use a subset of this package to fit only a Gaussian Process to the data. RadVel first performs a maximum-likelihood fit to the data and then determines errors through a Markov-Chain Monte Carlo (MCMC) analysis. We used 50 walkers, 2500000 steps, and a Gelman-Rubin statistic of 1.01 for convergence; the rest of the parameters are set to the default values as described in \citet{Fulton2018}.

\section{Results} 

\subsection{Solar Temporal Variations using EMPIRE}
\label{sec:EMPIRE}

To examine the time variation of the solar insolation and its GP hyperparameters, 
we perform a Gaussian process fit using a quasi-periodic kernel on each year of data separately. An example fit for one year of EMPIRE data is shown in Figure~\ref{fig:empirefit}. A year was chosen as the timescale so that sufficient rotation periods would occur in each group to accurately determine the parameters from the Gaussian process fit while still being short enough to be a plausible baseline for stellar photometry observations. We acknowledge one of the limitations with this method is we are monitoring discrete changes in the hyperparameters between years instead of as a continuous change. 

\begin{figure*}[htp]
\includegraphics[trim=50 50 50 0,clip,width=1\textwidth]{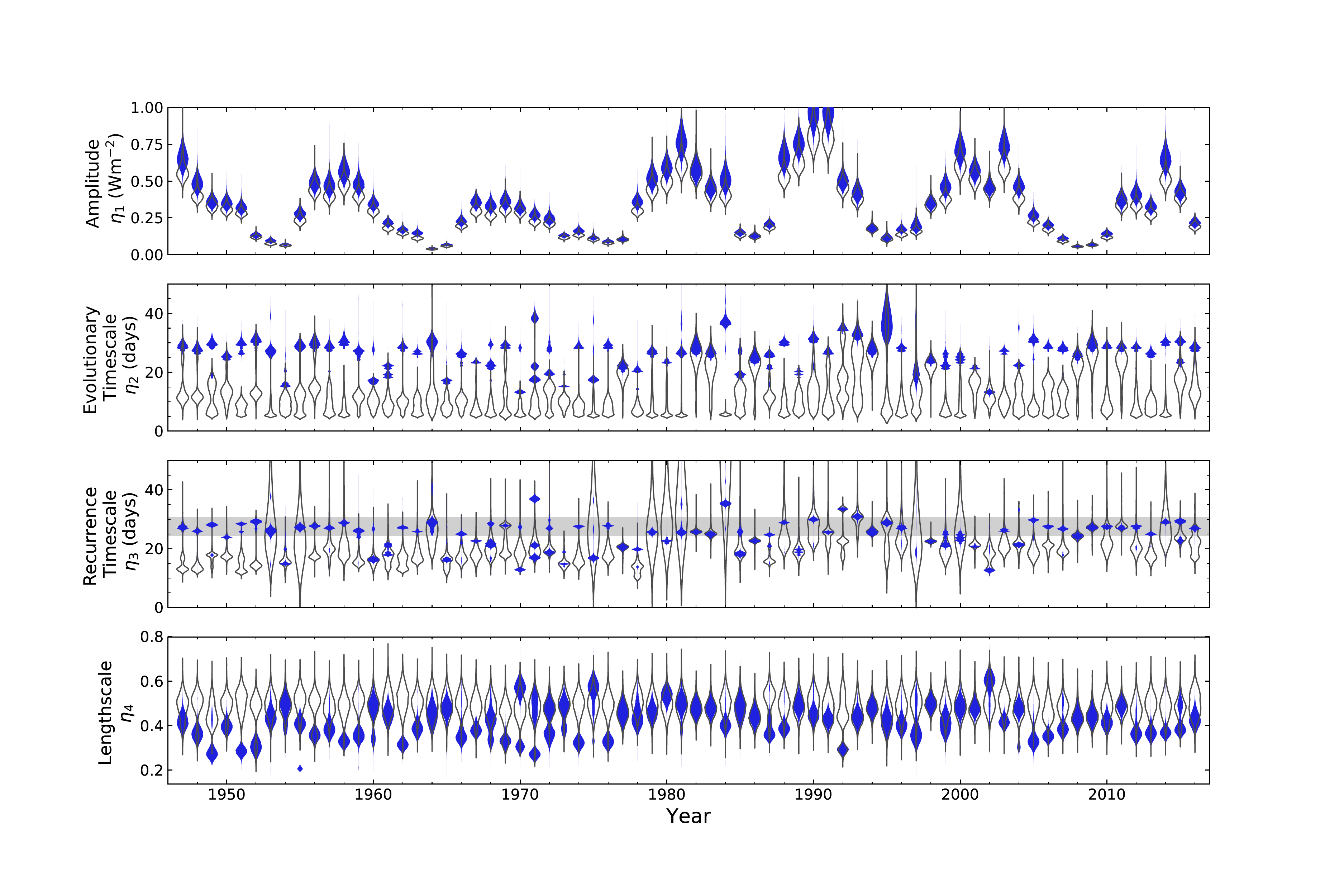}
\caption{Gaussian process hyperparameter posteriors for individual fits of each year of EMPIRE photometry, with (blue shaded) and without (black outline) a prior restricting the evolutionary timescale to be larger than the recurrence timescale. 
This added prior
results in a good match between the solar rotation period (shaded grey bar) and the recurrence timescale posterior. \label{fig:empiretime}}
\end{figure*}

We perform two fits for each year. The first fit has the following four priors. A non-informative prior is used on the amplitude (uniform prior of 0.01$<\eta_1<$10). As the data is sampled daily, small values for the evolutionary timescale allow the model to artificially change quickly enough to intersect all of the datapoints. Therefore, we limit the values of the evolutionary timescale on the lower end to avoid overfitting and the higher end as the model would be unable to detect a timescale longer than the data baseline (uniform prior of 5d$<\eta_2<$365d). For the recurrence timescale, we also limit the lower end to prevent overfitting and upper end at the baseline (uniform prior of 5d$<\eta_3<$365d). For other stars, the recurrence timescale can be constrained by a $vsin(i)$ measurement as short rotation periods produce large amplitudes or through determining the stellar rotation period through other methods.
Lastly, we constrain the length scale of the periodic component (Gaussian prior of $\eta_4$=0.5$\pm$0.05). The length scale is related to the average number of minima in a sample drawn from the Gaussian Process prior. An $\eta_4$ value of 0.5 means that there are on average two to three minima. \citet{Jeffers2009} finds that a random distribution of several active regions on the surface of a star produces two minima in the light curve, resulting in the Gaussian prior around 0.5 used in previous Gaussian process fits \citep[eg.][]{Haywood2014,LopezMorales2016}.

The posteriors of the four hyperparameters from 1947 to 2016 are shown in Figure~\ref{fig:empiretime}. The amplitude shows a clear 11 year variation matching the 11 year solar magnetic activity cycle. The variations also correlate well with the number of sunspots and inversely with the cosmic ray flux \citep{Usoskin2013}. 

The evolutionary timescale and recurrence timescale are interrelated. In years with an inferred low evolutionary timescale, the recurrence timescale is fairly unconstrained as the model is able to well-fit the data without a strong periodic component. The recurrence timescale describes the periodic component of the photometry and therefore should relate to the solar rotation period. The recurrence timescale posterior is well constrained at the solar rotation period for only a few years (1982, 1983, 1986, 1994, 2008, 2009, and 2011). These years all have something in common: the evolutionary timescale is longer than the recurrence timescale. For the majority of the other years, the inferred evolutionary timescale is shorter than the inferred rotation period. From this, it appears that the model is only successful in determining the rotation period if the evolutionary timescale is longer than the recurrence timescale. 


On the Sun, sunspot lifetime is proportional to the spot area \citep{Gnevyshev1938,Waldmeier1955}. Measured sunspot lifetimes range from a few days \citep{Petrovay1997} to hundreds of days \citep{Henwood2010}. 
However, the evolutionary timescale is not describing individual spot lifetimes but instead the evolution of large active regions. 
Measured lifetimes of solar active regions ranges from hours to months \citep{Schrijver2000,vanDriel-Gesztelyi2015}; the lifetime is roughly proportional to the active region's peak magnetic flux and can depend on the phase of the solar magnetic cycle and strength of surrounding magnetic fields.  Large active regions last from weeks to months, many of which have a longer lifetime than the solar rotation period, providing physical justification for a prior restricting the evolutionary timescale to be longer than the recurrence timescale. 

Furthermore, the timescales of active region evolution were estimated for 35 main sequence FGK stars through S-index measurements at Mount Willson Observatory \citep{Donahue1997}. The estimated lifetimes of these active regions ranged from 75 to 3000 days and the stellar rotation period ranged from 5 to 200 days with an average near 50 days. All of these stars have longer active region evolution timescales than their measured rotation periods, suggesting that this relationship holds for other FGK dwarf stars.

This relationship motivates our second fit where we include an additional prior to constrain the evolutionary timescale to be larger than the recurrence timescale ($\eta_2>\eta_3$). 
With this additional prior, the recurrence timescale is consistent with the solar rotation period to 1$\sigma$ for 48 of the 70 years. In addition, many of the previously multi-modal posteriors are now single peaks and the long tail posteriors are better constrained. 
If one is using a Gaussian process to determine a stellar rotation period, we recommend including this prior. 
The amplitude shows a small systematic increase with the additional prior; the trend with the solar magnetic cycle remains strong. The structure parameter has a greater variation between the years and has a lower average (approximately 0.4), favoring more high-frequency structure in the lightcurves.

\begin{figure*}[tpb]
\hspace*{-1cm}
\includegraphics[height=0.75\textwidth]{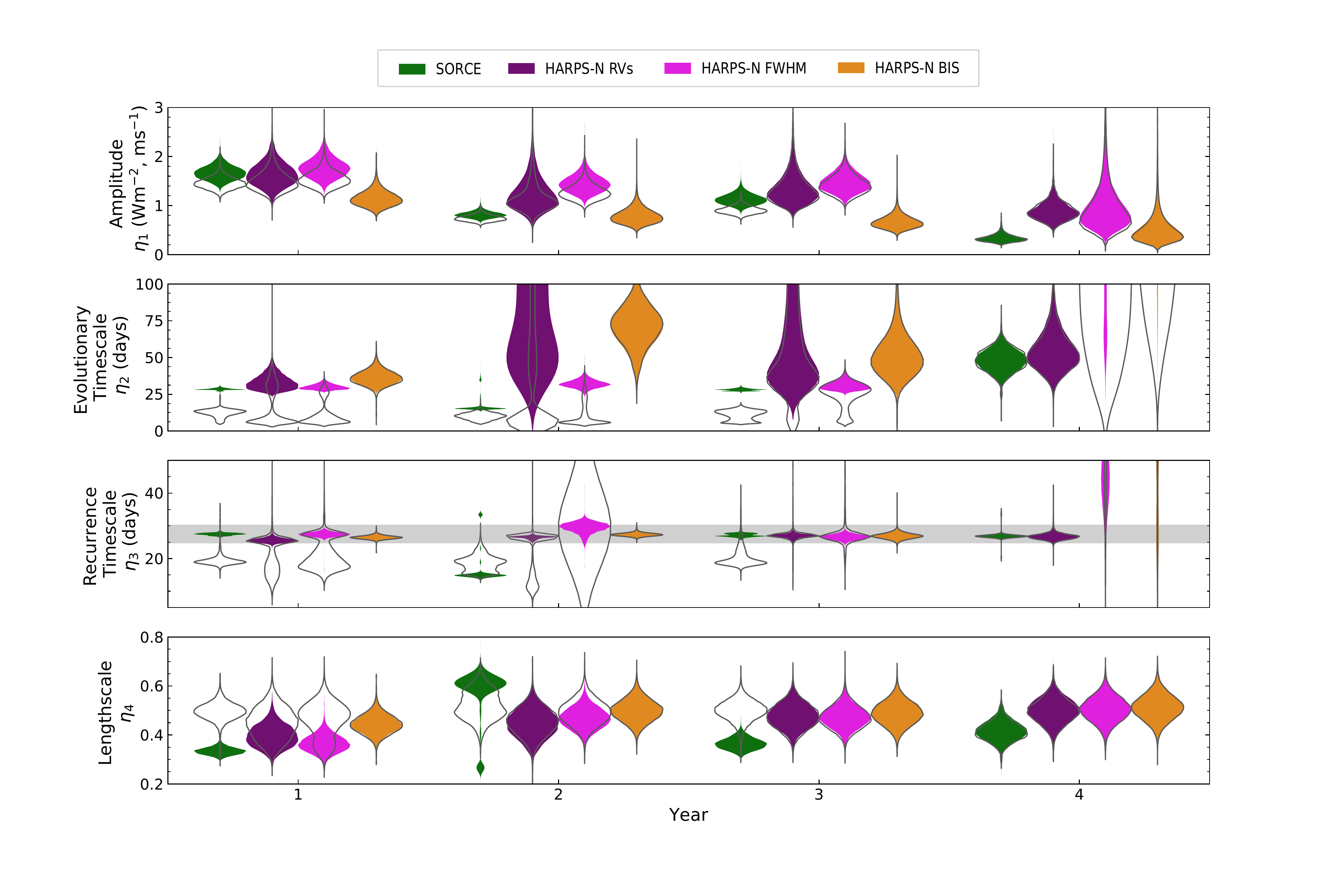}
\caption{Gaussian process hyperparameters for SORCE photometry, HARPS-N RVs, HARPS-N FWHM, and HARPS-N BIS for two fits, one with (shaded) and without (black outline) an additional prior constraining the evolutionary timescale to be longer than the recurrence timescale. The SORCE, FWHM, and BIS amplitudes have been scaled up by factors of 5, 1000, and 1000 respectively to be visible.
The recurrence timescales are largely consistent with the solar rotation period (shaded grey bar). The evolutionary timescale, recurrence timescale, and lengthscale posteriors are consistent between the four datasets for most years; therefore, photometry can provide valuable information about stellar activity through constraining these parameters for a radial velocity fit.}
\label{fig:comp_plot}
\end{figure*}

\subsection{Direct Comparison of Photometry with Radial Velocity Data}

Radial velocity data is often sparsely sampled and therefore poorly constrains the Gaussian process hyperparameters without additional information. In previous works, active stellar lines or photometry have been used to provide stellar activity information for the radial velocity fit \citep[eg.][]{Aigrain2012,Haywood2014,Kosiarek2019}. A key assumption in these analyses is that stellar activity is recorded in the same way between the two data types; however, for stars with low magnetic activity the radial velocity data may be dominated by phenomena not observable from a lightcurve \citep{Wright2005,Tayar2019}.  The overlap between the SORCE photometry dataset with the well-sampled HARPS-N radial velocity dataset provides an unique opportunity to test this assumption for sun-like stars.


The same procedure described above for the EMPIRE analysis (Section~\ref{sec:EMPIRE}) is performed here, and the results are shown in Figure~\ref{fig:comp_plot}. To recap, two fits are run for each dataset; the first (posteriors shown as a black outline) with the following two uniform priors: 5d$<\eta_2<$365d, 5d$<\eta_3<$365d and one Gaussian prior: $\eta_4$=0.5$\pm$0.05. The second (posteriors shown as a solid color interior) has an additional prior constraining the decay timescale to be larger than the recurrence timescale ($\eta_2>\eta_3$).

The main takeaway from these fits is that the posteriors are largely consistent between all datasets 
in the Gaussian process fit with the additional prior ($\eta_2>\eta_3$); therefore, photometry can provide valuable information about stellar activity for radial velocity analyses. In both fits, the amplitude posteriors are largely consistent within each dataset with a slight downward trend as the data approaches the solar minimum.

There are two interesting comparisons from the initial fit without the additional prior. First, the SORCE photometry and HARPS-N RVs have consistent posteriors that match the solar rotation period only in Year 4 where the SORCE data has a longer evolutionary timescale than recurrence timescale. 
Second, the FWHM and BIS show opposite results to the photometry. The FWHM data well matches the RVs for Year 1-3 and not Year 4. The BIS posteriors for Year 1-3 are consistent with the solar rotation period and has a longer evolutionary timescale than recurrence timescale.

The posteriors of the second analysis display much higher agreement between the different datasets. The SORCE photometry and HARPS-N RVs are consistent for three of the four years; the inconsistent year, Year 2, SORCE instead has a recurrence timescale of half of the solar rotation period. The RVs and FWHM posteriors are now both well constrained and the recurrence timescale matches the solar rotation for Years 1-3. The BIS remains unchanged for Years 1-3 as the evolutionary timescale was already longer than the recurrence timescale and the recurrence timescale matched the solar rotation period. Lastly, $\eta_4$ may be underconstrained in the three HARPS-N datasets as the posteriors closely resemble the Gaussian prior on $\eta_4$. The lengthscale parameter for the SORCE photometry is around 0.4, lower than the lengthscale parameter for the three HARPS-N datasets, consistent with the lengthscales found in the EMPIRE analysis (Section~\ref{sec:EMPIRE}).

Year 4 is distinct as the FWHM and BIS do not have well constrained posteriors and do not match the photometry or RVs. Additionally, Year 4 is near the solar minimum and is the one year that the SORCE photometry matched the HARPS-N RVs without the additional prior; perhaps solar activity displays different characteristics in line measurements compared to photometry throughout the solar cycle. Further high-cadence radial velocity monitoring of the sun will be important to confirm many of the observations from this paper and potentially detect changes as a function of the solar cycle. 

\section{Conclusion}

We analysed simultaneous disk-integrated photometry and radial velocity data of the Sun in order to determine the useful limits of a combined analysis. We examined the periodicities of five simultaneous datasets, SORCE and EMPIRE photometry, HARPS-N radial velocity, and two HARPS-N line indicators: FWHM and BIS. 
The periodograms and autocorrelation plots often displayed power at the stellar rotation period and its harmonics; however, the stellar rotation period was not always the highest peak. In the 70 year EMPIRE dataset, the highest peak matched the solar rotation period 14.3\% and 21.4\% of the time for our periodogram and autocorrelation analysis respectively. We recommend exercising caution when using either of these methods to determine a stellar rotation period due to the large number of peaks at times unrelated to the solar rotation period.

A Gaussian process analysis of photometry can provide more reliable estimates of a star's rotation period. 
We used a Gaussian process to investigate the time variability of solar photometry through analysing 70 years of EMPIRE data. 
The time variability analysis determined that the Gaussian process amplitude hyperparameter followed the eleven year solar magnetic cycle. The evolutionary timescale and recurrence timescales remained relatively stable throughout and the recurrence timescale matched the solar rotation period when the additional prior constraining the evolutionary timescale to be greater than the recurrence timescale was included. Therefore, this Gaussian process analysis identified the correct solar rotation period more often than either the periodogram or autocorrelation analyses.

Photometry can also be a valuable tool for understanding stellar activity in radial velocity data fits. In our direct comparisons between the Gaussian process hyperparameters of the SORCE photometry, HARPS-N RV data, and HARPS-N FWHM and BIS line measurements, the evolutionary timescale and recurrence timescale were consistent between the datasets after including the same additional prior restricting the evolutionary timescale to be longer than the recurrence timescale. 
We recommend including this additional prior to improve the agreement between Gaussian Process hyperparameters derived from photometry and radial velocity data.  
The lengthscale parameter was consistent between the four datasets, although the value for the photometry data was systematically low compared to the other three.  

Precision Radial Velocity surveys are aiming to characterize Earth-like planets around Solar-type stars with cm s$^{-1}$ radial velocity signals.  
Overlapping data spanning a full solar cycle or a few solar cycles is necessary to confirm the findings in this paper and to look for evidence for changes as a function of the solar cycle. Further work is also needed to determine how these conclusions could be applied to other stellar types.

%
%



\acknowledgments
Acknowledgements: 
The authors thank our anonymous reviewer for constructive comments that improved the clarity of this work. We also thank the California Planet Search team for helpful discussions that improved the quality of this work. 
M.R.K is supported by the NSF Graduate Research Fellowship, grant No. DGE 1339067.


Facilities:  \facility{SORCE} \facility{EMPIRE} \facility{TNG:HARPS-N}\\
Software: RadVel, Python

\bibliographystyle{aasjournal.bst}
\bibliography{main.bib}

\end{document}